\documentclass{article}
\usepackage{psfrag}
\usepackage{float}
\usepackage{epsf}
\usepackage{epsfig}
\usepackage{psfig}
\usepackage{fancyhdr}
\usepackage{latexsym} 
\usepackage{graphics}
\usepackage{graphicx}
\usepackage{amsfonts}
\usepackage{amsmath}

\newcommand{\beq}{\begin{equation}}
\newcommand{\eeq}{\end{equation}}
\newcommand{\beqa}{\begin{eqnarray}}
\newcommand{\eeqa}{\end{eqnarray}}

\begin{document}

\thispagestyle{empty}

\par
\topmargin=-1cm      

{ \small

\hfill \noindent{\tiny HISKP-TH-04/11}

\hfill \noindent{\tiny FZJ-IKP(TH)-2004-08}

}

\thispagestyle{empty}

\vspace{80.0pt}

\begin{centering}
{\Large\bf Orthonormalization procedure for chiral\\[0.3em]
 effective nuclear field theory}\footnote{Work
supported in part by Deutsche Forschungsgemeinschaft under contract
no. GL 87/34-1 and by funds provided from
the EU to the project ``Study of Strongly Interacting Matter'' under contract no.
RII3-CT-2004-506078.
}\\

\vspace{30.0pt}
{\bf H.~Krebs}$^1$,
{\bf V.~Bernard}$^2$,
{\bf Ulf-G.~Mei{\ss}ner}$^{1,3}$\\
\vspace{20.0pt}

{\sl $^{1}$Helmholtz-Institut f\"ur Strahlen- und Kernphysik (Theorie), 
Universit\"at Bonn}\\
{\sl D-53115 Bonn,
Germany} \\
{\it E-mail addresses: hkrebs@itkp.uni-bonn.de, meissner@itkp.uni-bonn.de}\\

\vspace{15.0pt}

{\sl $^{2}$Laboratoire de Physique Th\'eorique,
Universit\'e Louis Pasteur} \\
{\sl  F-67037 Strasbourg Cedex 2, France} \\
{\it E-mail address: bernard@lpt6.u-strasbg.fr}\\

\vspace{15.0pt}

{\sl $^{3}$Institut f\"ur Kernphysik (Theorie), Forschungszentrum J\"ulich}\\
{\sl D-52425 J\"ulich,
Germany} \\

\end{centering}
\vspace{20.0pt}
\begin{center}
\begin{abstract}
\noindent
We show that the $\hat Q$-box expansion of nuclear many-body physics can
be applied in nuclear effective field theory with explicit pions and
external sources. We establish  the corresponding power counting and give
an algorithm for the construction of a hermitean and energy-independent
potential for arbitrary scattering processes on nucleons and nuclei to a
given order in the chiral expansion. Various examples are discussed in some
detail.
\end{abstract}

\vspace*{50pt}
{\small
%
}
\vfill
\end{center}

\newpage


\section{Introduction and summary}
Effective field theory methods applied to few-nucleon systems have become
a powerful tool to establish a systematic and precise formulation of nuclear
physics consistent with the symmetries of QCD, in particular its spontaneously and
explicitly broken chiral symmetry. The field was initiated by a series of papers of
Weinberg \cite{Wein1,Wein2,Wein3} and has matured considerably. Recently, two-nucleon
dynamics based on effective field theory (EFT) with explicit pions~\footnote{We do not
consider the so-called pionless EFT in what follows.} has been worked
out at next-to-next-to-next-to-leading order~(NNNLO), see \cite{EM,EGM3}~\footnote{Note that 
only in \cite{EGM3} a detailed analysis of the theoretical uncertainties at this order 
is given as it is required for any sensible EFT analysis.}. Due to the appearance of
shallow bound states a purely perturbative treatment like in chiral
perturbation theory for
systems with pions or with one nucleon is inappropriate. However, as 
stressed already by Weinberg,
one can apply the power counting in the construction of the so-called 
effective potential.
Through iteration of this effective potential one obtains the full physical 
scattering amplitude. In the case of nucleon-nucleon
scattering, it was shown in \cite{Wein1} that the power counting is violated if 
one considers
diagrams containing two-nucleon intermediate states. From that, the simplest 
definition of the
effective potential can be given as follows: the effective potential is the sum of all 
time-ordered diagrams which contain in each intermediate state at least one pion. 
No violation
of the power counting appears on the level of the effective potential, thus it can 
be constructed
perturbatively in terms of the usual small parameters $q/\Lambda_\chi$ and 
$M_\pi/\Lambda_\chi$,
with $q$ a generic external momentum, $M_\pi$ the pion mass and $\Lambda_\chi
\simeq 1\,$GeV the scale of chiral
symmetry breaking. The extension of these arguments to pion production or the 
inclusion of the
$\Delta (1232)$ is straightforward if one takes into account the new scales
appearing in these cases (see e.g. \cite{CH}). We will not 
consider  such extensions here specifically, although our formalism is general
enough to include these. Once the effective potential is constructed to a
given order in the chiral expansion, the full $T$-matrix generating the bound
and scattering states can be calculated by solving the Lippmann-Schwinger
equation, $T = V_{\rm eff} +V_{\rm eff} G_0 T$, with $G_0$ the two-nucleon propagator.
Even though this definition of the effective potential is very simple, one
encounters some disadvantages. The potential is energy-dependent and not
hermitean. This makes it difficult to apply it to scattering processes involving
more than two nucleons. A solution to this problem was given by Epelbaum and
collaborators \cite{Epelbaum1}, who showed that the method of unitary 
transformations developed
by Fukuda, Sawada and Taketani \cite{Sawada} and by Okubo \cite{Okubo} can be applied to
generate an energy-independent and hermitean potential consistent with the
power counting of chiral perturbation theory  (the
effective potential). While this method is quite powerful, it suffers from the 
disadvantage that for different processes one has to define different model
spaces and each time to construct the operator $A$, which parameterizes the
corresponding unitary transformation, and to deduce the resulting hermitean
effective Hamiltonian. We propose here another scheme based on the 
${\hat Q}$-box expansion
of Kuo and collaborators \cite{Kuo1,Kuo2}. The main results of this study are:
\begin{itemize}
\item[i)] We have shown that the ${\hat Q}$-box expansion respects the chiral
expansion: only a finite number of ${\hat Q}$-boxes contribute to the effective
potential at a given chiral order.
\item[ii)] We have constructed an explicit algorithm that allows one to construct
the effective potential for any process involving nucleons, pions and photons 
(or other external sources) to a given order in the chiral expansion, 
see section~\ref{sec:Hefforder}.
\item[iii)] We have shown for various examples that when one requires the
on-shell condition to the asymptotic states, one recovers the expressions based
on time-ordered perturbation theory, as it should be.
\item[iv)] An explicit application of the method developed here is the calculation
of the fourth-order corrections to the three-body contributions in neutral
pion electroproduction off the deuteron, see \cite{HBMel}.
\end{itemize}

\medskip\noindent
The manuscript is organized as follows. In section~\ref{sec:Wein} we briefly review
Weinberg's approach to the chiral dynamics of few-nucleon systems and the corresponding
power counting. A general derivation of the hermitean effective potential is 
given in sec.~\ref{sec:hep}, following closely \cite{Suzuki1}. The ${\hat Q}$-box 
expansion is discussed in sec.~\ref{sec:qbox}. These three sections do not contain
new results but are needed to keep our presentation self-contained. 
Sec.~\ref{sec:Hefforder} contains our central results. We show that the ${\hat Q}$-box 
expansion is consistent with the chiral expansion and further give an explicit 
algorithm to construct the hermitean effective potential to a given chiral order.
Some leading order considerations are discussed in sec.~\ref{sec:leading}, in particular
the role of the so-called Okubo corrections (wave function orthonormalization
diagrams). Various technicalities are relegated to the appendices.

\section{A brief review of Weinberg's approach to chiral few-nucleon dynamics}
\label{sec:Wein}

This section contains a very brief introduction to chiral perturbation 
theory with two (or more) nucleons and serves mainly to fix the notation. 
As pointed out by 
Weinberg~\cite{Wein1,Wein2,Wein3} some years ago, for processes involving two 
or more nucleons
the standard power counting of chiral perturbation theory 
is valid only for a subset of diagrams. This subset consists of the
time-ordered diagrams  which in every intermediate state
contain at least one pion (or delta). The sum of all the diagrams from this 
subset is called the effective potential, denoted by $V_{\rm eff}^{NN}$. 
The whole transition matrix or nuclear 
wave function can be computed numerically in the standard way by
solving the inhomogeneous or homogeneous Lippmann-Schwinger equation with
the given effective potential. For a more detailed description, see 
e.g.~\cite{Epelbaum2}

\begin{figure}
\center
\includegraphics[height=2cm]{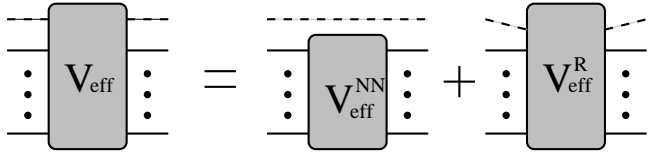}
\caption{A nuclear potential decomposed into the 
pion spectator and the interaction part.\label{piNNScat}} 
\end{figure}

If we are interested in pion scattering processes on nucleons or nuclei, we
can obviously proceed in a similar way and define an effective potential
composed of diagrams, which contain in every intermediate state at least
two pions (or deltas). In this way we get an effective potential matrix
\beq
V_{\rm eff}=\left(
\begin{array}{rr}
\langle\pi NN|V_{\rm eff}|\pi NN\rangle & \langle\pi NN|V_{\rm eff}|NN\rangle\\
\langle NN|V_{\rm eff}|\pi NN\rangle & \langle NN|V_{\rm eff}| NN\rangle 
\end{array}
\right),
\eeq 
where $NN$ denote here a state with two or more than two nucleons. 
The elastic pion
scattering potential can be decomposed in two parts as has been shown in 
Fig.~\ref{piNNScat}: the first part contains all diagrams from the effective
potential with the pion acting as a spectator. Let us denote it by 
$\langle\pi NN|V_{\rm eff}^S|\pi NN\rangle$.
 The second part contains all other
possible diagrams. It is obvious that the first part is given by the usual 
effective potential $V_{\rm eff}^{NN}$ multiplied by the three dimensional delta 
function of the pion spectator. If we want to describe  scattering 
processes on nuclei, it is useful to define the well--known 
Lovelace operator~\cite{Joachain}, called here the irreducible kernel $K$:
\beq\label{defK}
K=V_{\rm eff}^i+V_{\rm eff}^f\frac{1}{E-H_0-V_{\rm eff}+i\epsilon}
V_{\rm eff}^i,
\eeq
where $E$ denotes the total energy of the system and 
the potentials $V_{\rm eff}^i$ and $V_{\rm eff}^f$ are given by
\beq
V_{\rm eff}^i=H_0+V_{\rm eff}-H_0^i,\quad V_{\rm eff}^f=H_0+V_{\rm eff}-H_0^f,
\eeq
and $H_0$, $H_0^i$ and $H_0^f$ denote the free Hamilton operators of the particles in 
the intermediate state, of the initial and of the final asymptotic states, respectively. 
In the case of
pion scattering, e.g., $V_{\rm eff}^i$ and $V_{\rm eff}^f$
are given by the effective potential without the pion spectator contribution: 
\beq
V_{\rm eff}^i=V_{\rm eff}^f=V_{\rm eff}-
\left(
\begin{array}{cc}
\langle\pi NN|V_{\rm eff}^S|\pi NN\rangle & 0\\
0 & 0
\end{array}
\right) \, .
\eeq
All diagrams from $K$ are
called irreducible and all others are called reducible.
To calculate the transition amplitude for scattering processes
on nuclei, one has to convolute the irreducible kernel $K$ with the 
corresponding
nuclear wave functions. Since in every 
intermediate state of the kernel $K$ the chiral counting can be 
applied~(no infrared singularities appear if the nucleon mass approaches
infinity), one can calculate it perturbatively. 
The  simultaneous expansion both in small momenta (masses) of
Goldstone-bosons and in one over the nucleon mass can be performed.
The $1/m$ Taylor series expansion within the potentials can be done, since 
no unitarity cuts can appear there (note that we always define the
model space such that the potentials are real). The cuts,
which are responsible for the imaginary part of the scattering amplitude, 
are all outside the potentials (the inside/outside refers to the structure
of the kernel given in Eq.~(\ref{defK})), 
where the trivial $1/m$ Taylor expansion does not make
sense. However outside the potentials one can perform the $1/m$ expansion
within the energy denominator, which is symbolically given in the following
equation:
\beq
\frac{1}{E-E'+i \epsilon}=\frac{1}{(E-E')_0+A/m+B/m^2+\dots+i\epsilon}.
\eeq
Here $(E-E')_0$ denotes the corresponding energy difference $E-E'$ in the 
static limit. In some cases the nucleon recoil corrections, denoted here
by $A/m$, are sizeable and can not be neglected, for more detailed discussion 
of
this topic see e.g.~\cite{Nrecoil}.
The power counting for a given diagram from the irreducible kernel $K$ 
or nuclear potential is given by~\cite{Wein3}
\beq\label{powerc}
\nu=4-N-2C+2L+\sum_i V_i\Delta_i,\quad\Delta_i=d_i+\frac{1}{2}n_i-2,
\eeq
where $N$, $C$, $L$, $V_i$, $d_i$ and $n_i$ are the number of nucleons,
of connected pieces, of loops, of vertices of type $i$, 
of derivatives (or powers of the pion mass) at the vertex $i$ and 
of nucleons at the vertex $i$, respectively. In order to calculate
a scattering process up to a given chiral order $\nu_{max}$ we have to sum
up all the diagrams with $\nu\le\nu_{max}$ from the kernel $K$ and convolute
them with the chiral nuclear wave functions fixed by the corresponding
nuclear potential calculated up to order $\nu_{max}+3$. The added $3$ here
guarantees that the operators $V_{\rm eff}^S$ and $K$ are of the same chiral 
order~\footnote{Note that $V_{\rm eff}^S$ is given by a nuclear potential
multiplied by the pion spectator delta function, which decreases the chiral
dimension by $-3$.}. 

Although all that looks nice, one well known disadvantage appears: 
the effective
potentials defined in this simple way are energy-dependent and non-hermitean.
This makes it difficult to apply the formalism to the processes, where more
than two nucleons are involved. Epelbaoum et al.~\cite{Epelbaum1} suggested
another construction of effective potential using the techniques of unitary
transformation, firstly introduced by Okubo~\cite{Okubo} many years ago. 
They showed explicitly, how to construct the hermitean energy-independent 
$N$-nucleon potential up to a given order in chiral 
perturbation theory. To be consistent with this picture we have to extend 
the given construction
to the scattering processes on nucleons and nuclei. The main purpose of this
manuscript is to give such an extension, such that both the nuclear 
wave functions on the one hand and the irreducible kernel on the other hand
were constructed using hermitean energy-independent potentials 
in a unified way.

\section{Hermitean effective potential}
\label{sec:hep}

To derive the general effective potential, we proceed in the manner of
Suzuki and Okamoto \cite{Suzuki1}. For earlier and related work, see e.g.
\cite{Levy,Brandow,Suzuki0,Suzuki2,Suzuki3,Suzuki94,Suzuki4}.
Let the full Fock-space $\cal H$ of all
physical 
states be described as a direct sum of a model subspace $\cal H_M$ and 
its complementary space $\cal H_R$,
\begin{eqnarray}
\cal{H} &=& \cal{H_M}\oplus\cal{H_R}.
\end{eqnarray}  
Let $P$ and $Q$ be projection operators onto these two subspaces,
\begin{eqnarray}
P,Q:\cal{H} &\rightarrow& \cal{H}
\end{eqnarray}
such that $Im\lbrack P \rbrack\subset \cal{H_M}$, 
$Im\lbrack Q \rbrack\subset\cal{H_R}$, $P Q = Q P = 0$ and $P + Q = 1$, 
with $1=Id$ the Identity operator
$Id:\cal{H}\rightarrow\cal{H}$. Let $A :\cal{H}\rightarrow\cal{H}$ be 
an operator with the following property, 
$A\lbrack\cal{H_M}\rbrack\subset\cal{H_R}$ and 
$A\left[\cal{H_R}\right]=0$, such that it maps 
the states from the model space $\cal{H_M}$ to the states from the 
complementary space $\cal{H_R}$. From this property it follows immediately
\begin{eqnarray}
A &=& Q A P~,\\
A Q &=& 0~,\\
P A &=& 0~,
\end{eqnarray}
and
\begin{eqnarray}
A^2&=&0~.
\end{eqnarray} 
Now we define an operator $X(n)$ by
\begin{eqnarray}
X(n)&=&(1+A)(1+A^\dagger A+A A^\dagger)^n,
\end{eqnarray}
with $n$ a real number. The inverse of $X(n)$ is 
\begin{eqnarray}
X^{-1}(n)&=&(1+A^\dagger A+A A^\dagger)^{-n}(1-A).
\end{eqnarray}
The above relation can be verified by using the fact that 
$(1-A)(1+A)=1$. The stationary Schr\"odinger equation of the 
original problem $H\Psi=E\Psi$, where $H = H_0 + V$ is the Hamiltonian
of the physical problem represented as a sum of the free and
interacting parts $H_0$ and $V$, can be transformed to
\begin{eqnarray}
X^{-1}(n)H X(n)X^{-1}(n)\Psi&=&E X^{-1}(n)\Psi.
\end{eqnarray} 
Suppose that the transformed operator $X^{-1}(n)H X(n)$ fulfills for
$n=0$ the decoupling equation
\begin{eqnarray}
\label{decouple}
Q X^{-1}(0)H X(0)P&=&Q(H + \lbrack H ,A\rbrack - A H A)P \,=\,0.
\end{eqnarray}
This is exactly the equation of Okubo \cite{Okubo} for the operator $A$.
The above equation leads to the general decoupling equation
\begin{eqnarray}
Q X^{-1}(n)H X(n)P&=&(Q+A A^\dagger)^{-n}Q X^{-1}(0)H X(0)P 
(P+A^\dagger A)^n\,=\,0,
\end{eqnarray}
where we have used the fact, that
\begin{eqnarray}
\label{PProjAA}
(1+A^\dagger A+A A^\dagger)^n P&=&(P+A^\dagger A)^n,
\\
Q(1+A^\dagger A+A A^\dagger)^n&=&(Q+A A^\dagger)^n.
\end{eqnarray}
If we now start from the effective equation in the model space 
\begin{eqnarray}
P X^{-1}(n)H X(n)P \Phi&=& E \Phi,\,\,\,\,\,\, \Phi\in\cal{H_M}
\end{eqnarray}
we get (because of the decoupling equation)
\begin{eqnarray}
X^{-1}(n)H X(n) \Phi&=& E \Phi.
\end{eqnarray}
Multiplying this equation by $X(n)$ from the left brings us to the
original Schr\"odinger equation:
\begin{eqnarray}
H \Psi&=& E \Psi,
\end{eqnarray}
with $\Psi=X(n)\Phi$.
Let us introduce at this stage some important notation for the effective 
Hamiltonian and the potential:
\begin{eqnarray}
H_{\rm eff}(n)&:=&P X^{-1}(n)H X(n)P\,=P H_0 P+V_{\rm eff}(n)
\end{eqnarray}
To study the properties of $V_{\rm eff}(n)$, we 
multiply the Eq.~$(\ref{decouple})$ from the left by $A^\dagger$
and conjugate the last equation:
\begin{eqnarray}
P A^\dagger Q H P + P A^\dagger Q H Q A P 
- P A^\dagger A 
P H P - P A^\dagger A P H Q A P&=&0,\\
P H QA P + PA^\dagger Q H Q A P - P H P A^\dagger A 
P - P A^\dagger Q H P A^\dagger A P &=&0.
\end{eqnarray}  
Eliminating the $QHQ$-term from the above equations we get
\begin{eqnarray}
P(1+A^\dagger)H P(P+A^\dagger A)&=&(P+A^\dagger A)P H(1+A)P.
\end{eqnarray}
The effective Hamiltonian and its adjoint have the form
\begin{eqnarray}
H_{\rm eff}(n)&=&(P+A^\dagger A)^{-n}(1-A)H(1+A)(P+A^\dagger A)^n\\
&=&(P+A^\dagger A)^{-n}PH(1+A)P(P+A^\dagger A)^n\nonumber,\\
H_{\rm eff}^\dagger(n)&=&(P+A^\dagger A)^{n}(1+A^\dagger)H(1-A^\dagger)(P+A^\dagger A)^{-n}\\
&=&(P+A^\dagger A)^{n}P(1+A^\dagger)H P(P+A^\dagger A)
(P+A^\dagger A)^{-n-1}\nonumber\\
&=&(P+A^\dagger A)^{n+1}P H (1+A)
(P+A^\dagger A)^{-n-1}.\nonumber
\end{eqnarray}
So we obtain the relation
\begin{eqnarray}
H_{\rm eff}(n)&=&H_{\rm eff}^\dagger(-n-1).
\end{eqnarray}
From the above relation we see that $H_{\rm eff}(n)$ becomes hermitean 
when $n=-1/2$.

For practical purposes two effective interactions are of special 
importance. The first one is $H_{\rm eff}(0)$, which has a relatively simple form
\begin{eqnarray}
\label{EnergieElim}
H_{\rm eff}(0)&=& P H P + P H Q A\,=\,P H_0 P + R,
\end{eqnarray}
with the non-hermitean potential, denoted by 
\begin{eqnarray}
\label{RPotential}
R = V_{\rm eff}(0)\,=\,P V P + P V Q A.
\end{eqnarray}
Because of its simplicity this energy-independent non-hermitean potential 
has rich applications in nuclear physics.

\medskip\noindent
Another important effective Hamiltonian is $H_{\rm eff}(-1/2)$, which is hermitean.
Its explicit form is
\begin{eqnarray}
H_{\rm eff}(-1/2)&=&(P+A^\dagger A)^{1/2}H(P+A)
(P+A^\dagger A)^{-1/2}\,=\,P H_0 P + W,
\end{eqnarray}
with the hermitean potential, denoted by $W = V_{\rm eff}(-1/2)$.
The hermitean theory is more complicated than the standard non-hermitean
theory, but has an advantage that the wave functions, which come from the
solution of the effective Schr\"odinger equation, are orthonormal. 
To see this, let us define an operator
\beq
G={\rm arctanh}\left(A-A^\dagger\right).
\eeq
It can be shown that the following relation is valid~\cite{Suzuki2}:
\beq
\exp(G)=\left(1+A-A^\dagger\right)
\left(1+A^\dagger A+A A^\dagger\right)^{-1/2}.
\eeq
Using Eq.~(\ref{PProjAA}) we immediately see that
\beq
\exp(G)P=\left(1+A\right)\left(P+A^\dagger A\right)^{-1/2}=X(-1/2)P.
\eeq
Let $\{|\psi_n\rangle\}$ denote a set of orthonormal eigenstates
of the Hamiltonian $H$. The corresponding eigenstates $\{|\phi_n\rangle\}$ 
of effective Hamiltonian are related to the original one by
\beq
|\psi_n\rangle=X(-1/2)|\phi_n\rangle=X(-1/2)P|\phi_n\rangle=
\exp(G)P|\phi_n\rangle,
\eeq 
such that we get for the effective eigenstates
\beq
|\phi_n\rangle=\exp(-G)|\psi_n\rangle.
\eeq 
The orthonormality of the effective eigenstates follows now directly from the
unitarity of $\exp(G)$:
\beq
\langle\phi_n|\phi_m\rangle=\langle\psi_n|\exp(G)\exp(-G)|\psi_m\rangle=
\langle\psi_n|\psi_m\rangle=\delta_{n,m}.
\eeq

\medskip\noindent
Now let us discuss some interesting relations between hermitean and non-hermitean
potentials. One can show the following general relation \cite{Suzuki0}
\begin{eqnarray}
W&=&P V P + \sum_{m,n=0}^{\infty}F(m,n)\left((A^\dagger A)^m P V Q A
(A^\dagger A)^n + h.c.\right),
\end{eqnarray}
with the function $F(m,n)$ defined as coefficients of the Taylor series of
\begin{eqnarray}
f(x,y)&=&\frac{\sqrt{1+x}}{\sqrt{1+x}+\sqrt{1+y}}\,=\,\sum_{m,n=0}^{\infty}F(m,n)x^m y^n.
\end{eqnarray}
From the property $f(x,y)+f(y,x)=1$ we have the anti-symmetry relation
\begin{eqnarray}
\label{AntiSymmFmn}
F(m,n)&=&-F(n,m),
\end{eqnarray}
unless $m=n=0$. The values of $F(m,n)$ for small $m$ and $n$ are given in
Table~\ref{Fmn}.
\begin{table}
\caption{The values of $F(m,n)$.}\label{Fmn}
\begin{center}
\begin{tabular}{|c|c|c|c|c|c|c|}
\hline
$(m,n)$  &$(0,0)$ &$(0,1)$ &$(1,0)$ &$(0,2)$ &$(1,1)$ &$(2,0)$\\
\hline 
$F(m,n)$ &$1/2$ &$-1/8$ &$1/8$ &$1/16$ &$0$ &$-1/16$\\
\hline
\end{tabular}
\end{center}
\end{table}
Due to the anti-symmetry relation $(\ref{AntiSymmFmn})$ and from the definition
of $R$ in Eq.~$(\ref{RPotential})$, we can write $W$ in another form as
\begin{eqnarray}
\label{WPotGeneral}
W&=&\frac{R+R^\dagger}{2}+\sum_{m+n > 0}F(m,n)(A^\dagger A)^m
(R-R^\dagger)(A^\dagger A)^n,
\end{eqnarray} 
where the summation runs over zero or all positive integers except $m=n=0$.
The expression of $W$ in the above equation clarifies how the non-hermiticity
term
$\sim(R-R^\dagger)$ contributes to the hermitean effective interaction $W$.
The first two terms in the expansion of $W$ are given by
\begin{eqnarray}
\label{VeffReduced}
W&=&\frac{1}{2}(R + R^\dagger)+\frac{1}{8}\left((A^\dagger A)
(R-R^\dagger) + h.c.\right) - \frac{1}{16}\left((A^\dagger A)^2
(R-R^\dagger) + h.c.\right) + \ldots~.
\end{eqnarray}
This formula will be useful for our later investigations.

\section{The $\hat{Q}$-box expansion of the effective interaction}
\label{sec:qbox}

\begin{figure}
\center
\includegraphics[height=5cm]{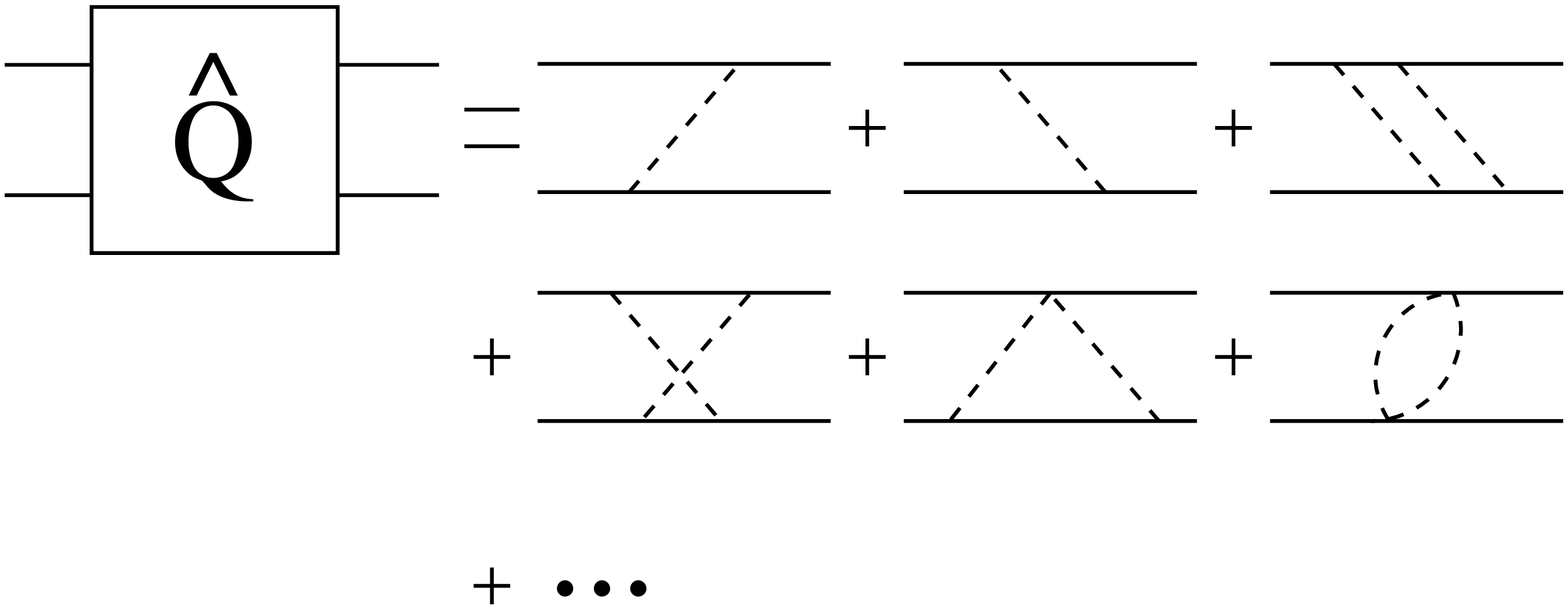}
\caption{Some lowest order diagrams of a $\hat{Q}$-box in the case
of nucleon-nucleon scattering.\label{qboxNN}} 
\end{figure}

We are now interested in the solution of the decoupling equation 
$(\ref{decouple})$ for the operator $A$. Once we have constructed
the operator $A$ we are able to give an explicit form for the
effective potentials $R$ and $W$. For this reason we define a so called
$\hat{Q}$-box, which was first introduced by Kuo and collaborators\cite{Kuo1} 
in the study of the folded diagram expansion of an effective potential:
\begin{eqnarray}
\hat{Q}(E)&=&P V P + P V Q\frac{1}{E - Q H Q}Q V P.
\end{eqnarray} 
The physical meaning of this object is very familiar. It is nothing else
but the sum of all time-ordered diagrams, which do not contain any 
model-space intermediate states.
For example if one fixes the model-space as 
a space of all pure nucleon states, then every diagram from the $\hat{Q}$-box
has to contain at least one pion or delta in every intermediate state. Some
lowest order diagrams of a $\hat{Q}$-box in the case of nucleon-nucleon 
scattering are given in Figure~\ref{qboxNN}.
We also want to introduce higher $\hat{Q}$-boxes as
\begin{eqnarray}
\hat{Q}_n(E_1,...,E_{n+1})&=&(-1)^{n}P V Q\frac{1}
{(E_1-Q H Q)\cdots(E_{n+1}-Q H Q)}Q V P.
\end{eqnarray}
Higher $\hat{Q}$-boxes can be easily derived from the original $\hat{Q}$-box
through the relation \cite{Suzuki94}
\begin{eqnarray}
\label{PartDecomp}
\hat{Q}_n(E_1,...,E_{n+1})&=&\sum_{k=1}^{n+1}C_k(E_1,...,E_{n+1})\hat{Q}(E_k)
\end{eqnarray}
where
\begin{eqnarray}
\label{CKonstK}
C_k(E_1,...,E_{n+1})&=&\frac{1}{(E_k-E_1)\cdots(E_k-E_{k-1})(E_k-E_{k+1})\cdots
(E_k-E_{n+1})}.\,\,\,\,\,\,\,\,\,\,\,\,\,\,\,
\end{eqnarray}
If at least two of the energies mentioned above are equal, 
one has to understand 
Eq.~$(\ref{PartDecomp})$ as a limit. To make this clear, let us look for 
example at $\hat{Q}_1(E_1,E_2)$-box, with the energies $E_1=E_2$:
\begin{eqnarray}
&\lim_{E_1\rightarrow E_2}&\left(
\frac{1}{\left(E_1-E_2\right)}\hat{Q}\left(E_1\right)
+\frac{1}{\left(E_2-E_1\right)}\hat{Q}\left(E_2\right)
\right)
\nonumber\\
=&\lim_{E_1\rightarrow E_2}&\left(
\frac{1}{\left(E_1-E_2\right)}P V P
+\frac{1}{\left(E_2-E_1\right)}P V P
\right)\nonumber\\
+&\lim_{E_1\rightarrow E_2}&P V Q\frac{1}{E_1-E_2}\left(
\frac{1}{E_1-Q H Q}-\frac{1}{E_2-Q H Q}
\right)Q V P
\nonumber\\
=&\lim_{E_1\rightarrow E_2}&P V Q\frac{1}{E_1-E_2}\left(
\frac{E_2-E_1}{\left(E_1-Q H Q\right)\left(E_2-Q H Q\right)}\right)Q V P
\nonumber\\
=&-&P V Q\frac{1}{(E_2-Q H Q)^2}Q V P.\nonumber
\end{eqnarray}
With these definitions one can express the potentials $R$ and $W$ in terms of
$\hat{Q}$--boxes. 
Let us classify all contributions to $W$ by a  $\kappa$ number given by
\beq\label{defkappa}
\kappa\left(\hat{Q}_{k_1}\dots\hat{Q}_{k_n}\right)=k_1+\dots+k_n.
\eeq
Denote by $W_n$ all $\hat{Q}$-box contributions to $W$ with $\kappa=n$.
The potential can be expressed as an infinite sum of those contributions:
\beq
W=\sum_{n=0}^{\infty} W_n.
\eeq  
Here we want to derive explicit expressions for $W_0$, $W_1$ and $W_2$.
For this reason let $|\alpha\rangle$ be an eigenstate 
of $P H_0 P$ with an energy $E_\alpha$
from the model space, then the decoupling equation $(\ref{decouple})$ can
be written as
\begin{eqnarray}
\label{decoupleGen}
A|\alpha\rangle&=&\frac{1}{E_\alpha-Q H Q}Q V P|\alpha\rangle - 
\frac{1}{E_\alpha-Q H Q}A R|\alpha\rangle.
\end{eqnarray}
For brevity let us now adopt a notation that whenever a symbol 
$|\beta\rangle\langle\beta|$ appears the summation over $\beta$ has to be
understood. Now we eliminate $A$ from the
potential $R$:
\beqa
R|\alpha\rangle&=&P V P|\alpha\rangle + P V Q A|\alpha\rangle\nonumber\\
&=&\hat{Q}(E_\alpha)|\alpha\rangle-P V Q\frac{1}{E_\alpha-Q H Q}
A|\beta\rangle\langle\beta| R|\alpha\rangle\nonumber
\eeqa
\beqa
&=&\hat{Q}(E_\alpha)|\alpha\rangle+\hat{Q}_1(E_\alpha,E_\beta)
|\beta\rangle\langle\beta|R|\alpha\rangle
+ P V Q \frac{1}{E_\alpha-Q H Q}\frac{1}{E_\beta-Q H Q}A
|\gamma\rangle\langle\gamma| R
|\beta\rangle\langle\beta| R|\alpha\rangle\\\nonumber
&=&\hat{Q}(E_\alpha)|\alpha\rangle+\hat{Q}_1(E_\alpha,E_\beta)
|\beta\rangle\langle\beta|R|\alpha\rangle
+ \hat{Q}_2(E_\alpha,E_\beta,E_\gamma)
|\gamma\rangle\langle\gamma|R|\beta\rangle\langle\beta|R|\alpha\rangle.
\eeqa
An iterative solution of the previous equation can be given by
\beqa
R|\alpha\rangle&=&\hat{Q}(E_\alpha)|\alpha\rangle+\hat{Q}_1(E_\alpha,E_\beta)
|\beta\rangle\langle\beta|\hat{Q}(E_\alpha)|\alpha\rangle\nonumber\\
&+&\hat{Q}_1(E_\alpha,E_\beta)
|\beta\rangle\langle\beta|\hat{Q}_1(E_\alpha,E_\gamma)
|\gamma\rangle\langle\gamma|\hat{Q}(E_\alpha)|\alpha\rangle\\
&+&\hat{Q}_2(E_\alpha,E_\beta,E_\gamma)|\gamma\rangle\langle\gamma|
\hat{Q}(E_\beta)|\beta\rangle\langle\beta|\hat{Q}(E_\alpha)|\alpha\rangle
+\dots.\nonumber
\eeqa
To derive an approximate equation for the operator $A$ we
implement the previous equation for $R$ in Eq.~$(\ref{decoupleGen})$ and get
\begin{eqnarray}
A|\alpha\rangle&=&\frac{1}{E_\alpha-Q H Q}Q V P|\alpha\rangle - 
\frac{1}{E_\alpha-Q H Q}A|\beta\rangle\langle\beta|
\hat{Q}(E_\alpha)|\alpha\rangle,
\end{eqnarray}
where we only take the first term of $R$ , because we are not interested
in the calculation of $A$, but of $A^\dagger A$ (the neglected terms
of  $R$ would lead to contributions with $\kappa>2$). The iterative
solution of the above equation is given by
\begin{eqnarray}
A|\alpha\rangle&=&\frac{1}{E_\alpha-Q H Q}Q V P|\alpha\rangle
- \frac{1}{E_\alpha-Q H Q}\frac{1}{E_\beta-Q H Q}Q V P
|\beta\rangle\langle\beta|\hat{Q}(E_\alpha)|\alpha\rangle.
\end{eqnarray}
For the operator $A^\dagger A$ we get
\begin{eqnarray}
\label{WOp}
\langle\delta|A^\dagger A|\alpha\rangle&=& - \langle\delta|
\hat{Q}_1(E_\delta,E_\alpha)|\alpha\rangle - \langle\delta|
\hat{Q}_2(E_\alpha,E_\beta,E_\delta)|\beta\rangle\langle\beta|\hat{Q}(E_\alpha)
|\alpha\rangle\\
&-&\langle\delta|\hat{Q}(E_\delta)|\beta\rangle\langle\beta|
\hat{Q}_2(E_\alpha,E_\beta,E_\delta)|\alpha\rangle.\nonumber
\end{eqnarray}
If we now use the expressions for the operators
$R$ and $A^\dagger A$ in Eq.~$(\ref{VeffReduced})$ (note that the
$\hat Q$-box expansion of $A^\dagger A$ starts with a $\hat Q_1$-box such 
that all terms of Eq.(\ref{WPotGeneral}) with $m+n > 2$ lead to contributions 
with $\kappa > 2$.),  we get
the following explicit form of $W_0$, $W_1$ and $W_2$: 
\beqa
\langle\delta|W_0|\alpha\rangle
&=&\frac{1}{2}\biggl\{\langle\delta|\hat{Q}(E_\alpha)|\alpha\rangle
+\langle\delta|\hat{Q}(E_\delta)|\alpha\rangle\biggr\},\label{WPotHermit1}\\
\langle\delta|W_1|\alpha\rangle
&=&\frac{1}{2}\biggl\{\langle\delta|\hat{Q}_1(E_\alpha,E_\beta)
|\beta\rangle\langle\beta|\hat{Q}(E_\alpha)|\alpha\rangle
+\langle\delta|\hat{Q}(E_\delta)|\beta\rangle\langle\beta|
\hat{Q}_1(E_\beta,E_\delta)|\alpha\rangle\biggr\}\nonumber\\
&+&\frac{1}{8}\biggl\{- \langle\delta|
\hat{Q}_1(E_\delta,E_\beta)|\beta\rangle\left(\langle\beta|\hat{Q}(E_\alpha)
|\alpha\rangle - \langle\beta|\hat{Q}(E_\beta)
|\alpha\rangle\right )\nonumber\\
&-&\left(\langle\delta|\hat{Q}(E_\delta)
|\beta\rangle - \langle\delta|\hat{Q}(E_\beta)
|\beta\rangle\right)\langle\beta|
\hat{Q}_1(E_\beta,E_\alpha)|\alpha\rangle\biggr\},\label{WPotHermit2}\\
\langle\delta|W_2|\alpha\rangle
&=&\frac{1}{2}\biggl\{\langle\delta|\hat{Q}_1(E_\alpha,E_\beta)
|\beta\rangle\langle\beta|\hat{Q}_1(E_\alpha,E_\gamma)
|\gamma\rangle\langle\gamma|\hat{Q}(E_\alpha)|\alpha\rangle
+\langle\delta|\hat{Q}(E_\delta)
|\gamma\rangle\langle\gamma|\hat{Q}_1(E_\delta,E_\gamma)
|\beta\rangle\langle\beta|\hat{Q}_1(E_\delta,E_\beta)|\alpha\rangle\nonumber\\
&+&\langle\delta|\hat{Q}_2(E_\alpha,E_\beta,E_\gamma)|\gamma\rangle\langle\gamma|
\hat{Q}(E_\beta)|\beta\rangle\langle\beta|\hat{Q}(E_\alpha)
|\alpha\rangle
+\langle\delta|\hat{Q}(E_\delta)|\beta\rangle\langle\beta|
\hat{Q}(E_\beta)|\gamma\rangle\langle\gamma|\hat{Q}_2(E_\delta,E_\beta,E_\gamma)
|\alpha\rangle\biggr\}\nonumber\\
&+&\frac{1}{8}\biggl\{
- \langle\delta|
\hat{Q}_1(E_\delta,E_\beta)|\beta\rangle
\left(\langle\beta|
\hat{Q}_1(E_\alpha,E_\gamma)
|\gamma\rangle\langle\gamma|\hat{Q}(E_\alpha)|\alpha\rangle
 - \langle\beta|\hat{Q}(E_\beta)
|\gamma\rangle\langle\gamma|\hat{Q}_1(E_\beta,E_\gamma)
|\alpha\rangle\right )\nonumber\\
&-&\left(\langle\delta|\hat{Q}(E_\delta)
|\beta\rangle\langle\beta|\hat{Q}_1(E_\delta,E_\beta)
|\gamma\rangle-\langle\delta|
\hat{Q}_1(E_\gamma,E_\beta)
|\beta\rangle\langle\beta|\hat{Q}(E_\gamma)|\gamma\rangle\right )
\langle\gamma|\hat{Q}_1(E_\gamma,E_\alpha)|\alpha\rangle\nonumber\\
&-&\langle\delta|\hat{Q}_2(E_\gamma,E_\beta,E_\delta)|\beta\rangle\langle\beta|
\hat{Q}(E_\gamma)|\gamma\rangle\left(\langle\gamma|\hat{Q}(E_\alpha)
|\alpha\rangle-\langle\gamma|\hat{Q}(E_\gamma)|\alpha\rangle\right)\nonumber\\
&-&\left(\langle\delta|\hat{Q}(E_\delta)|\beta\rangle-
\langle\delta|\hat{Q}(E_\beta)|\beta\rangle\right)\langle\beta|
\hat{Q}_2(E_\alpha,E\gamma,E_\beta)|\gamma\rangle\langle\gamma|\hat{Q}(E_\alpha)
|\alpha\rangle\nonumber\\
&-&\langle\delta|\hat{Q}(E_\delta)|\beta\rangle\langle\beta|
\hat{Q}_2(E_\gamma,E_\beta,E_\delta)|\gamma\rangle
\left(\langle\gamma|\hat{Q}(E_\alpha)
|\alpha\rangle-\langle\gamma|\hat{Q}(E_\gamma)|\alpha\rangle\right)\nonumber\\
&-&\left(\langle\delta|\hat{Q}(E_\delta)|\beta\rangle-
\langle\delta|\hat{Q}(E_\beta)|\beta\rangle\right)\langle\beta|
\hat{Q}(E_\beta)|\gamma\rangle\langle\gamma|\hat{Q}_2(E_\alpha,E_\gamma,E_\beta)
|\alpha\rangle\biggr\}\nonumber\\
&-&\frac{1}{16}\biggr\{\langle\delta|\hat{Q}_1(E_\delta,E_\beta)
|\beta\rangle\langle\beta|\hat{Q}_1(E_\beta,E_\gamma)|\gamma\rangle
\left(\langle\gamma|\hat{Q}(E_\alpha)|\alpha\rangle - 
\langle\gamma|\hat{Q}(E_\gamma)|\alpha\rangle\right)\nonumber\\
&+&\left(\langle\delta|\hat{Q}(E_\delta)|\beta\rangle - \langle\delta|
\hat{Q}(E_\beta)|\beta\rangle\right)
\langle\beta|\hat{Q}_1(E_\beta,E_\gamma)|\gamma\rangle\langle\gamma|
\hat{Q}_1(E_\gamma,E_\alpha)|\alpha\rangle\biggr\}.\label{WPotHermit3}
\eeqa
Until now we did not specify the model-space, so the above equation is 
quite general and can be applied to every problem in chiral perturbation
theory with more than one nucleon.

\section{Hermitean potential up to a given order}
\label{sec:Hefforder}

The main advantage of the $\hat{Q}$-box expansion is that it respects the 
chiral expansion: only a finite number of $\hat{Q}$-boxes contribute to the 
effective potential up to a given chiral order. 

\begin{figure}
\center
\includegraphics[height=2.5cm]{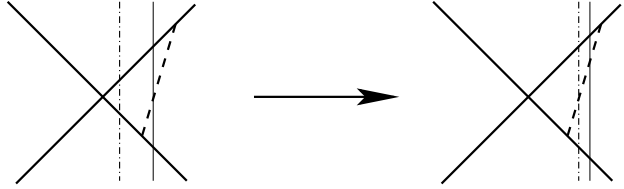}
\caption{Triangle nucleon--nucleon diagram. The ordinary time-ordered reducible
diagram on the left-hand-side becomes an irreducible one by shifting 
the left (dot-dashed) cut. One gets an irreducible diagram on the right-hand-side
with one insertion from a $\hat{Q}_1$-box.\label{trian}} 
\end{figure}

To demonstrate this issue, 
we fix the model space as consisting of states with two or more nucleons. The
states with at least one pion or delta are assumed to belong to the 
complementary space.
Now we can give a diagrammatical interpretation of Eqs.~$(\ref{WPotHermit1}),
(\ref{WPotHermit2})$ and $(\ref{WPotHermit3})$ 
considering
nucleon-nucleon scattering. The interpretation of Eq.~$(\ref{WPotHermit1})$
is already given in Fig.~\ref{qboxNN}. 
To give an interpretation of the other two Eqs.~$(\ref{WPotHermit2})$ and $(\ref{WPotHermit3})$ consider first the triangle diagram of Fig.~\ref{trian}. 
By shifting the $NN$ cut to the $NN\pi$ cut we get an irreducible diagram
with one insertion from a $\hat{Q}_1$-box such that this diagram contributes
to the term  $\langle\delta|\hat{Q}(E_\delta)|\beta\rangle\langle\beta|
\hat{Q}_1(E_\beta,E_\delta)|\alpha\rangle$. Even more all possible 
contributions to $\langle\delta|\hat{Q}(E_\delta)|\beta\rangle\langle\beta|
\hat{Q}_1(E_\beta,E_\delta)|\alpha\rangle$  are generated in
the same way:
every contribution of it can be visualized by a definite diagram with exactly 
one $NN$ intermediate cut which has to be shifted to the complementary space intermediate
states in every possible way. The same is valid for terms including higher
$\hat{Q}$-boxes: all contributions of terms with $\kappa=n$ can be visualized
by definite diagrams with exactly $n$ $NN$ intermediate cuts, which have to
be shifted to the complementary space intermediate states in every possible way.

From this interpretation we see that the higher the $\kappa$-number of
a $\hat{Q}$-box contribution to the potential $W$ is, the higher is the number 
of $NN$ cuts which have to be shifted and consequently the higher is the
number of loops which increases the chiral order of a given contribution. For
this reason there is only a finite number of $\hat{Q}$-box contributions if
the chiral order is fixed. Although we restricted here our considerations
to nucleon-nucleon scattering, the last statement is quite general: we
could choose other model and complementary spaces and show with the same arguments
that the $\hat{Q}$-box expansion of the effective potential $W$ stops after
a finite number of terms if the chiral order is fixed~\footnote{A more formal 
proof for pion scattering processes off nucleons is presented in the appendix}.

\begin{figure}
\center
\includegraphics[height=2.5cm]{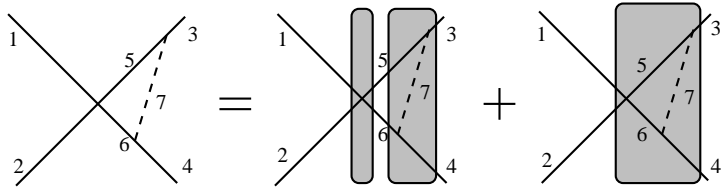}
\caption{One-loop triangle nucleon-nucleon time-ordered diagram written in 
terms of hermitean potentials denoted by gray shaded area.
\label{trianRec}} 
\end{figure}

Now we are at the point of showing how one can construct the effective 
potential $W$ up to given chiral order diagrammatically. The algorithm is indeed
very simple:
\begin{itemize}
\item Draw all possible covariant diagrams~(reducible and irreducible), 
which contribute up to a given ``naive''~\footnote{By ``naive'' we mean
here that we do not take into account the violation of the power counting by 
the reducible diagrams.} chiral order. 
\item Draw for every covariant diagram all possible time-orderings
and classify them by numbers~($\kappa$-numbers) of model space intermediate 
cuts.
\item In each time-ordered diagram with $\kappa=n$ shift all $n$ model 
space intermediate
cuts to the complementary space cuts in every possible way. Every diagram generated
in this way represents a contribution to a definite $\hat{Q}$-box 
structure with $\kappa=n$. Summing up all these contributions to the energy 
propagator we get the irreducible energy denominator of a given time-ordered
diagram.
\item The energy denominator of a given covariant diagram is then given by 
the sum of contributions from all possible time-orderings.
\item If $\kappa>0$ and there are no possibilities to shift the model space
intermediate cuts, then the corresponding diagram does not contribute to
the effective potential $W$.
\end{itemize}
The diagrams constructed according to this procedure obey the power counting, cf.
Eq.(\ref{powerc}).
To make the procedure clear let us discuss some examples:
\begin{itemize}
\item The first one is the triangle diagram in nucleon-nucleon 
scattering shown in Fig.~\ref{CovTrian}. The model space in this case is generated by
pure nucleonic states and the complementary space by states with at least one pion
or delta. There are only two possible time-ordered diagrams with $\kappa=1$.
The only possibility for shifting the nucleon-nucleon cut is to shift it
to the $\pi NN$-state (see Fig.~\ref{CovTrian}). All possible contributions
to the potential $W$ with $\kappa=1$  are given in  Eq.~$(\ref{WPotHermit2})$.
In this case the states $|\alpha\rangle$, $|\beta\rangle$ and 
$|\delta\rangle$ denote the two nucleon states $|N_3N_4\rangle$, 
$|N_5N_6\rangle$ and $|N_1N_2\rangle$, respectively. Consequently 
\beq
E_\alpha=E_3+E_4,\quad E_\beta=E_5+E_6,\quad E_\delta=E_1+E_2.
\eeq 
The energy denominators for the upper and lower time-ordered diagrams are given
by
\beq
P_u=-\frac{1}{2}\frac{1}{E_1+E_2-E_5-E_7-E_4}\frac{1}{E_5+E_6-E_5-E_7-E_4}
\eeq
and
\beq
P_l=-\frac{1}{2}\frac{1}{E_1+E_2-E_3-E_7-E_6}\frac{1}{E_5+E_6-E_3-E_7-E_6},
\eeq
respectively. The contribution weighted by $1/8$ from Eq.~$(\ref{WPotHermit2})$
vanishes because the corresponding $\hat{Q}$-box difference is of vertex type:
\beq
\langle\beta|\hat{Q}(E_\alpha)|\alpha\rangle
-\langle\beta|\hat{Q}(E_\beta)|\alpha\rangle=
\langle\beta|V|\alpha\rangle-\langle\beta|V|\alpha\rangle+\dots.\nonumber,
\eeq
where $\langle\beta|V|\alpha\rangle$ denotes the NN vertex.
Ellipses denote here terms of higher order, which don't contribute to the
considered diagram.
The 
energy denominator $P$ for the covariant triangle diagram is then given by
\beq
P=P_u+P_l.
\eeq
\begin{figure}
\center
\includegraphics[height=4.5cm]{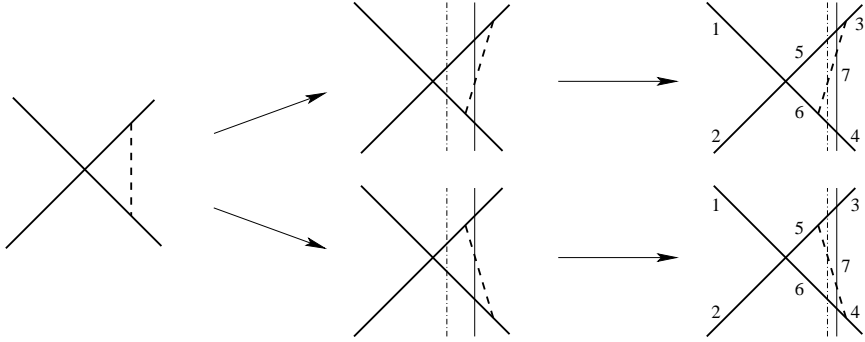}
\caption{Triangle nucleon-nucleon diagram of chiral order  $\nu=2$.
The two steps needed to determine the energy denominator are shown. 
For the discussion see the main text.
\label{CovTrian}} 
\end{figure}
\item As the second example let us consider one-loop pion production
diagram shown in Fig.~\ref{pionPrTr}. The model space in this case is generated
by pure nucleonic states and nucleonic states with one pion or delta. 
The states including two or more pions or deltas belong now to the complementary space.
As we can see from Fig.~\ref{pionPrTr} there are again two time-orderings. The
difference to the previous case is that in the upper diagram there are no
complementary space intermediate cuts. For this reason the model-space-cuts of the
upper diagram can not be shifted and consequently this time-ordered diagram
does not contribute to the potential matrix $W$. In the lower diagram there
is a two-nucleon-model-space-cut which can be shifted to the two-pion 
two-nucleon complementary space cut. The only contribution to the energy 
propagator comes from this diagram. To give the energy denominator we can use
Eq.~$(\ref{WPotHermit2})$, since the $\kappa$ number of the remaining diagram 
is equal one:
\beq
P=-\frac{1}{2}\frac{1}{E_1+E_2-E_5-E_3-E_8-E_7}\frac{1}{E_6+E_7-E_5-E_3-E_8
-E_7}.
\eeq 
\begin{figure}
\center
\includegraphics[height=4.5cm]{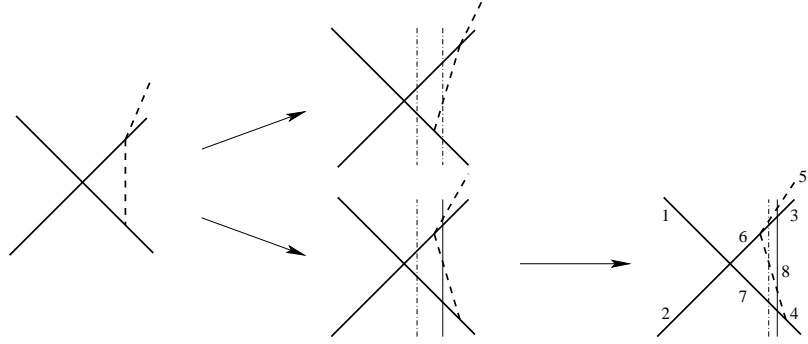}
\caption{One-loop pion production diagram of chiral order  $\nu=2$.
There are two possible time-orderings, which generate 1 diagram by shifting
the $NN$ cut to the cut with two pions.
\label{pionPrTr}} 
\end{figure}
\item As the third example let us consider pion production one-loop 
diagram shown in Fig.~\ref{pionPr}. In this case there are three time-orderings
with $\kappa=1$ and one time-ordering with $\kappa=2$. By shifting the model
space intermediate cuts we get 6 diagrams with $\kappa=1$ and one diagram
with $\kappa=2$. To give the energy denominators of the first 6 diagrams
with $\kappa=1$ we can use Eq.~$(\ref{WPotHermit2})$:
\beqa
P_1&=&-\frac{1}{2}\frac{1}{E_1+E_2-E_1-E_6-E_7-E_4}
\frac{1}{E_1+E_2-E_5-E_9-E_7-E_4}\nonumber\\
&\times&\frac{1}{E_8+E_7+E_4-E_5-E_9-E_7-E_4}\nonumber\\
&+&\frac{1}{8}\frac{1}{E_8+E_7+E_4-E_5-E_9-E_7-E_4}
\frac{1}{E_5+E_3+E_4-E_5-E_9-E_7-E_4}\nonumber\\
&\times&\left(\frac{1}{E_1+E_2-E_1-E_6-E_7-E_4}-
\frac{1}{E_8+E_7+E_4-E_1-E_6-E_7-E_4}\right),\nonumber\\
P_2&=&-\frac{1}{2}\frac{1}{E_5+E_3+E_4-E_1-E_6-E_7-E_4}
\frac{1}{E_8+E_7+E_4-E_1-E_6-E_7-E_4}\nonumber\\
&\times&\frac{1}{E_5+E_3+E_4-E_5-E_9-E_7-E_4}\nonumber\\
&+&\frac{1}{8}\frac{1}{E_1+E_2-E_1-E_6-E_7-E_4}
\frac{1}{E_8+E_7+E_4-E_1-E_6-E_7-E_4}\nonumber\\
&\times&\left(\frac{1}{E_5+E_3+E_4-E_5-E_9-E_7-E_4}-
\frac{1}{E_8+E_7+E_4-E_5-E_9-E_7-E_4}\right),\nonumber\\
P_3&=&-\frac{1}{2}\frac{1}{E_1+E_2-E_5-E_9-E_6-E_2}
\frac{1}{E_1+E_2-E_5-E_9-E_7-E_4}\nonumber\\
&\times&\frac{1}{E_8+E_6+E_2-E_5-E_9-E_7-E_4}\nonumber\\
P_4&=&-\frac{1}{2}\frac{1}{E_1+E_2-E_5-E_9-E_6-E_2}
\frac{1}{E_8+E_6+E_2-E_5-E_9-E_6-E_2}\nonumber\\
&\times&\frac{1}{E_8+E_6+E_2-E_5-E_9-E_7-E_4}\nonumber\\
\eeqa
\beqa
P_5&=&-\frac{1}{2}\frac{1}{E_1+E_2-E_5-E_9-E_6-E_2}
\frac{1}{E_8+E_6+E_2-E_5-E_9-E_6-E_2}\nonumber\\
&\times&\frac{1}{E_8+E_6+E_2-E_5-E_3-E_7-E_6-E_2}\nonumber\\
P_6&=&-\frac{1}{2}\frac{1}{E_1+E_2-E_5-E_9-E_6-E_2}
\frac{1}{E_1+E_2-E_5-E_3-E_7-E_6-E_2}\nonumber\\
&\times&\frac{1}{E_8+E_6+E_2-E_5-E_3-E_7-E_6-E_2}\nonumber
\eeqa
\beqa
P_7&=&\frac{1}{2}\frac{1}{E_1+E_2-E_5-E_9-E_7-E_4}
\frac{1}{E_8+E_6+E_2-E_5-E_9-E_7-E_4}\nonumber\\
&\times&\frac{1}{E_8+E_7+E_4-E_5-E_9-E_7-E_4}
\eeqa
The whole energy denominator is then given by a sum of all seven contributions:
\beq
P=P_1+\dots+P_7.
\eeq
To give the expression for $P_7$ we used Eq.~$(\ref{WPotHermit3})$ for 
the $\kappa=2$ contribution to $W$.
Note that there are no contributions to $P_7$ weighted by $1/8$ and $1/16$, 
since the corresponding $\hat{Q}$-box differences are of vertex type.
\begin{figure}
\center
\includegraphics[height=6.5cm]{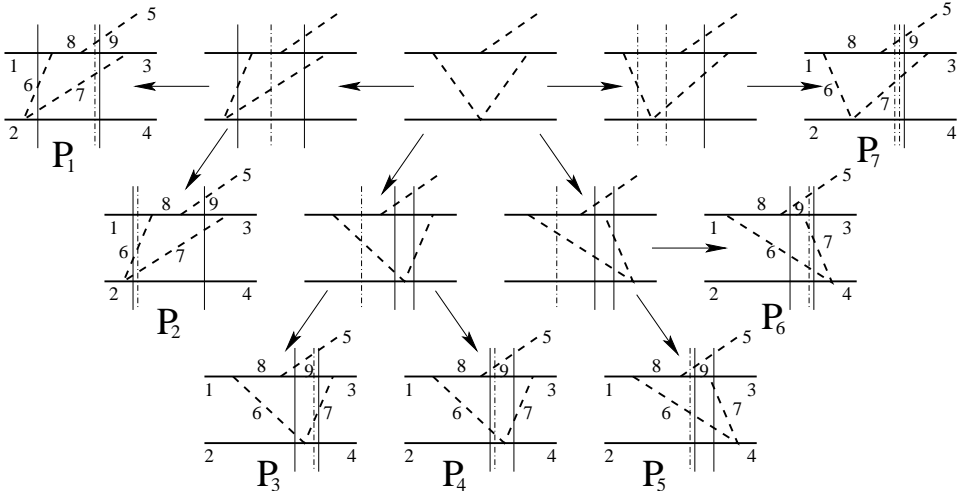}
\caption{One-loop pion production diagram of chiral order $\nu=2$ (in standard
Weinberg counting).
There are four different time-orderings, which generate 7 different diagrams
by shifting the $\pi NN$ intermediate cuts to the cuts with two or more pions. 
\label{pionPr}} 
\end{figure}
\end{itemize}
From the third example we see that the expressions for the energy denominators 
can be rather lengthy. For this reason it might be nice to have an independent
check of them. This indeed can be easily done: If one requires the on-shell
condition (the initial and final energy of a given scattering process are
required to be equal), then every time-ordered diagram~(reducible or 
irreducible) can be written in terms of hermitean potentials, 
including some iterations from the Lippman-Schwinger equation. 
This is illustrated in Fig.~\ref{trianRec} for the one-loop reducible triangle 
nucleon-nucleon scattering diagram. The corresponding
potentials are denoted by a gray shaded area. The energy denominators of 
both diagrams, the first one~(which comes from the first iteration in the 
Lippman-Schwinger equation) composed of two contributions and the
second one composed of one contribution from the hermitean potential, are given
by
\beq
T_1=\frac{1}{E-E_5-E_6}\frac{1}{2}\left(\frac{1}{E_5+E_6-E_5-E_7-E_4}+
\frac{1}{E-E_5-E_7-E_4}\right)
\eeq
and
\beq
T_2=-\frac{1}{2}\frac{1}{E-E_5-E_7-E_4}\frac{1}{E_5+E_6-E_5-E_7-E_4},
\eeq
respectively. Here we use the on shell condition $E=E_1+E_2=E_3+E_4$. If we 
now sum up the two contributions $T_1$ and $T_2$ we get the familiar
energy denominator of the corresponding time-ordered diagram\footnote{Note
that this calculation serves only to check the correctness of energy 
denominators in the effective potential. A strictly perturbative approach 
makes no sense in the case of nucleon-nucleon scattering.}:
\beqa
T_1+T_2&=&\frac{1}{E-E_5-E_6}\frac{1}{E-E_5-E_7-E_4}.
\eeqa
In the same way one should be able to reconstruct every possible 
time-ordered diagram. 

\section{Some leading order considerations}
\label{sec:leading}

\begin{figure}[htb]
\center
\includegraphics[height=3cm]{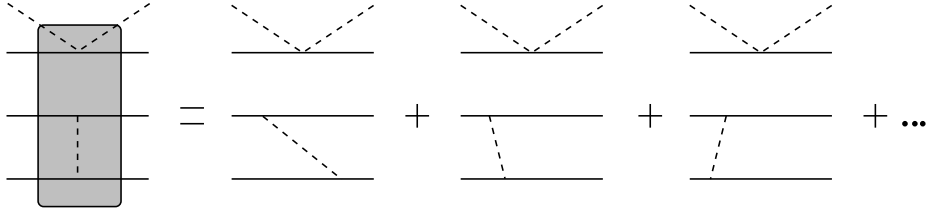}
\caption{Pion scattering on three nucleons: time-ordered diagram with 
two Okubo-type corrections. The ellipses denote the other time-ordered diagram
with the corresponding Okubo-type correction.
\label{time3Npi}} 
\end{figure}

\begin{figure}[htb]
\center
\includegraphics[height=2.5cm]{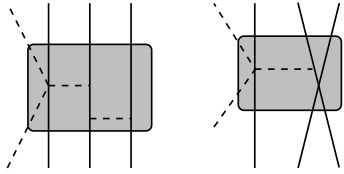}
\caption{Two diagrams that contribute to the hermitean potential in the case
of pion scattering off three nucleons.
\label{pi3nlikeNN}} 
\end{figure}

Let us briefly discuss the differences between some hermitean and
 non-hermitean potentials in the case of pion scattering on few-nucleon
systems. We denote the corresponding potentials here by $W$ and $W'$, respectively. 
Let us now consider some diagrams, which contribute to the potentials
$W$ or $W'$ at  leading order in the
$1/m$ expansion, namely in the static limit $m\rightarrow \infty$. 
If we assume that the incoming and
outgoing pions of the potentials $W$ or $W'$ are not involved in further 
interactions
in the kernel, then (because of the on-shell condition) their energies have to 
be equal in the static limit. In this case contributions to the potential
$W'$ can be identified with the $\kappa=0$ contributions to $W$. All 
other contributions to $W$ with $\kappa>0$ are called Okubo-type
corrections (these corrections are also called wave function
 orthonormalization diagrams \cite{EdenGari}).
\begin{figure}[hb]
\center
\includegraphics[height=2.4cm]{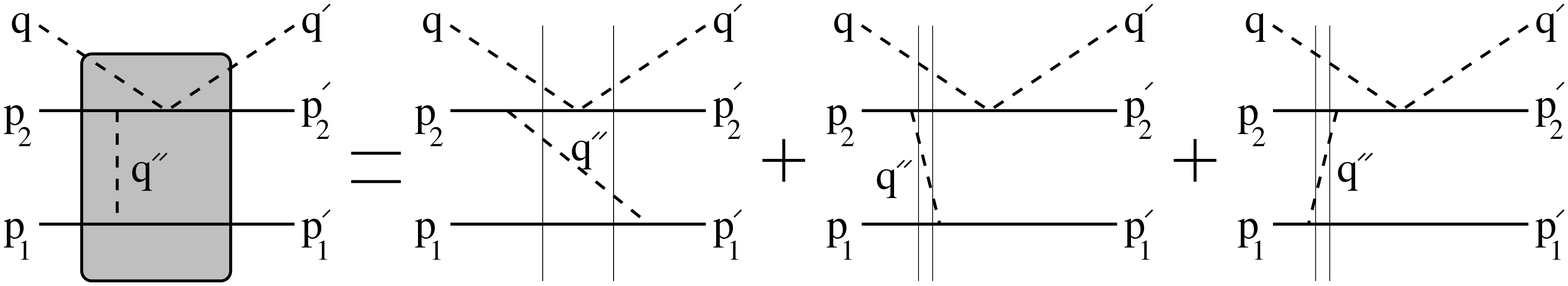}
\caption{Time-ordered diagram with two Okubo-type corrections for pion
scattering off two nucleons. 
\label{timepiNN}} 
\end{figure}
In some cases the Okubo-type corrections cancel some time-ordered diagrams from
$W'$. Consider e.g. the pion scattering diagrams shown in Fig.~\ref{timepiNN}.
The first time-ordered diagram contributes obviously to $W$ and $W'$.
The two other diagrams are Okubo-type corrections. 
The energy denominators of these three diagrams are given by 
\beq
P_1=\frac{1}{E_{q''}^2}~,\quad
P_2=-\frac{1}{2}\frac{1}{E_{q''}^2}~,\quad
P_3=-\frac{1}{2}\frac{1}{E_{q''}^2}~,
\eeq
such that their sum cancels. Here $E_{q''}$ denotes the energy of the 
rescattered pion. The same happens if we consider the pion-scattering
diagrams on three nucleons, shown in Fig.~\ref{time3Npi}. 
The Okubo-corrections cancel again the time-ordered diagram. Like in the
case of three nucleon scattering~\cite{Wein1,VanKolck1,EdenGari,Epelbaum1} 
it is straightforward to see 
that the contributions to $W$ in the case of pion scattering on three nucleons,
shown in Fig.~\ref{pi3nlikeNN}, vanish.

$\,$

\section*{Acknowledgements}

We are grateful to Evgeny Epelbaum for his numerous useful comments.

\bigskip
\appendix
\def\theequation{\Alph{section}.\arabic{equation}}
\setcounter{equation}{0}
\section{The degenerate case}
The aim of the three following appendices is to give the formal proof that 
the $\hat{Q}$-box contributions with $\kappa>2$ (see Eq.~(\ref{defkappa}))
lead in the case of pure nucleonic scattering to operators of chiral orders
 $\nu>10-3N$ and in the case of pion scattering off 
nucleons to  operators of chiral orders $\nu>7-3N$. 
Such kind of operators contribute in the case of pure 
nucleonic scattering to the orders higher than NNNLO. In the case of 
pion scattering on nucleons they lead to the kernel contributions of order 
higher than ${\cal O} (q^4)$ in the conventional counting, see e.g.
\cite{BBEPM} for a detailed discussion.

Let us assume for the moment that all states from the model-space have
the same energy $E$. In this case we can express
the potentials $R$ and $W$ in terms of $\hat{Q}$-boxes in a compact way.
The operator $P H_0 P$ can now be replaced by $E P$  and
the higher $\hat{Q}$-boxes by
\begin{eqnarray}
\hat{Q}_n&=&\frac{1}{n!}\frac{d^n\hat{Q}}{d E^n}\\
&=&(-1)^n P V Q\left(\frac{1}{E - Q H Q}\right)^{n+1}Q V P.\nonumber
\end{eqnarray}
The decoupling equation $(\ref{decouple})$ for the operator $A$ becomes
\begin{eqnarray}
\label{OmegaDegen}
A&=&\frac{1}{E-Q H Q}Q V P - \frac{1}{E-Q H Q}A R,
\end{eqnarray}
with $R$ as given in the Eq.~$(\ref{RPotential})$
\begin{eqnarray}
R&=&P V P + P V Q A.\nonumber
\end{eqnarray}
Eliminating $A$ from both equations above 
and using the definition of $\hat{Q}_n$
we obtain
\begin{eqnarray}
R&=&\sum_{n=0}^{\infty}\hat{Q}_n R^n,
\end{eqnarray}
where we have used the notation $\hat{Q}_0=\hat{Q}$ for $n=0$. This equation
can be solved exactly as \cite{Suzuki1}
\begin{eqnarray}
R&=&\sum_{k=1}^{\infty}\sum_{m_1,...,m_k}f(m_1,...,m_k)\hat{Q}_{m_1}\cdots
\hat{Q}_{m_k}.
\end{eqnarray}
The $f(m_1,...,m_k)$ are coefficients defined by
\begin{eqnarray}
\label{CoefFfunc}
f(m_1,...,m_k)&=&
\left\{\begin{array}{ll}
\delta_{m_1,0} & \textrm{for $k\,=\,1$} \\
\delta_{m_2,0}\delta_{m_1,1} &\textrm{for $k\,=\,2$}\\
\delta_{m_k,0}\delta_{m_1+\cdots+m_k,k-1}\prod_{i=1}^{k-2}\theta\left(i-\sum_{j=1}^{i} 
m_{k-j}\right) &\textrm{for $k\ge3$},
\end{array}\right.\,\,\,\,\,\,\,\,\,\,\,\,\,\,\,
\end{eqnarray}
where $\theta(x)$ is the step function
\begin{eqnarray}
\theta(x)&=&\left\{
\begin{array}{ll}
1 & \textrm{for $x\ge0$}\\
0 & \textrm{for $x<0$}.
\end{array}
\right.
\end{eqnarray}
Let us introduce a convenient notation for representing the series of
$R$. We define a symbol $\lbrack m\rbrack_k$, which denotes a set of k
elements which are zero or a positive integer
\begin{eqnarray}
\lbrack m\rbrack_k&=&(m_1,...,m_k).
\end{eqnarray}
We further define
\begin{eqnarray}
\hat{Q}\lbrack m\rbrack_k&=&\hat{Q}_{m_1}\cdots\hat{Q}_{m_k}.
\end{eqnarray}
Using these simple notations, we may write $R$ as
\begin{eqnarray}
R&=&\sum_{k=1}^{\infty}\sum_{\lbrack m \rbrack_k}f(\lbrack m \rbrack_k)\hat{Q}
\lbrack m \rbrack_k.
\end{eqnarray}
The coefficient $f(\lbrack m \rbrack_k)$ has an interesting property:
\begin{eqnarray}
f(\lbrack m \rbrack_{k_1})\cdots f(\lbrack m \rbrack_{k_n})&=&
f(n,\lbrack m \rbrack_{k_1},...,\lbrack m \rbrack_{k_n}),
\end{eqnarray}
such that the operator $R^n$ can be written as
\begin{eqnarray}
R^n&=&\sum_{k=0}^{\infty}\sum_{\lbrack m\rbrack_k}f(n,\lbrack m\rbrack_k)
\hat{Q}\lbrack m\rbrack_k,
\end{eqnarray}
where we use the notation
\begin{eqnarray}
f(n,\lbrack m \rbrack_0)&=&f(n).
\end{eqnarray}
Now we can solve Eq.~(\ref{OmegaDegen}) for $A$ exactly:
iterating Eq.~(\ref{OmegaDegen}) we have
\begin{eqnarray}
A&=&\sum_{n=0}^{\infty}(-1)^n\left(\frac{1}{E-Q H Q}\right)^{n+1}Q V P R^n\\
&=&\sum_{n=0}^{\infty}\sum_{k=0}^{\infty}\sum_{\lbrack m \rbrack_k}(-1)^n
f(n,\lbrack m \rbrack_k)\left(\frac{1}{E-Q H Q}\right)^{n+1}Q V P 
\hat{Q}\lbrack m \rbrack_k.
\end{eqnarray}
Multiplication with $A^\dagger$ gives
\begin{eqnarray}
\label{OmegaDaggerO}
A^\dagger A&=&-\sum_{p=0}^{\infty}\sum_{q=0}^{\infty}
\sum_{k=0}^{\infty}\sum_{l=0}^{\infty}\sum_{\lbrack m \rbrack_k}
\sum_{\lbrack n \rbrack_l}f(p,\lbrack m \rbrack_k)f(q,\lbrack n \rbrack_l)
\hat{Q}^\dagger\lbrack m \rbrack_k\hat{Q}_{p+q+1}\hat{Q}\lbrack n \rbrack_l.
\end{eqnarray}
Therefore, we can write $A^\dagger A$ as a sum of definite
combinations of $\hat{Q}$-boxes. We see from the above equation that the 
coefficients in front of the $\hat{Q}$-box combinations in the sum 
have the value
$0$ or $1$. They serve to exclude all impossible $\hat{Q}$-box
combinations. From the definition of the function $f$ in Eq.~(\ref{CoefFfunc})
we see, that the coefficients in the sum of Eq.~(\ref{OmegaDaggerO}) are zero
unless
\begin{eqnarray}
p+m_1+\cdots+m_k&=&k\quad\textrm{and}\quad q+n_1+\cdots+n_l\,=\,l
\end{eqnarray}
If we write $A^\dagger A$ as
\begin{eqnarray}
A^\dagger A&=&-\sum_{r=1}^{\infty}\sum_{\lbrack m \rbrack_r}
h(\lbrack m \rbrack_r)\hat{Q}\lbrack m \rbrack_r,
\end{eqnarray}
we obtain that the coefficients $h(\lbrack m \rbrack_r)$ are zero 
unless $m_1+\cdots+m_r=r$. This and the similar property of the coefficients
$f(\lbrack m \rbrack_r)$, namely they are zero unless $m_1+\cdots+m_{r-1}=r-1$
and $m_r=0$, will play an important role for the later power counting arguments.

\section{Power counting for $\hat Q$-boxes}
Let us derive the power counting for the $\hat{Q}(E)$-box following
Weinberg's arguments.  A general
time-ordered diagram of the $\hat{Q}$-box has the following form:
\begin{eqnarray}
\int (d^3q)^L\frac{1}{(E_k-E_l)^R}\left(\frac{1}{\sqrt{2 E_\pi}}\right)^{2I_p}\prod_i q^{d_iV_i}\nonumber
\end{eqnarray}
where $L$ is the number of loops, $R$ is the number of energy-denominators,
$d_i$ is the dimension of a vertex $i$ and  $V_i$ counts how often 
the vertex of 
type $i$ appears in the diagram. Under $I_p$, $E_p$, $I_n$ and $E_n$ we 
understand the number of inner pion, external pion, inner nucleon and external 
nucleon-lines. For the chiral order $\nu$ one gets
\begin{eqnarray}
\nu&=&3 L - R - I_p + \sum_{i}V_i d_i - 3 (C + D_p + D_n - 1),
\end{eqnarray}
where $C$, $D_p$ and $D_n$ denote the number of connected, disconnected pion
(pion spectators) and disconnected nucleon (nucleon spectators) pieces, 
respectively. 
Each connected and disconnected piece brings
a delta function, which carries dimension $-3$. The overall delta function
is not taken into account, for this reason $(C + D_p + D_n -1)$ appears in the 
above equation. The phase-factors $1/\sqrt{2E_\pi}$ of external pions are
also not taken into account. 
With the help of the topological identities
\begin{eqnarray}
\sum_{i}V_in_i&=&2 I_n + E_n - 2 D_n,\\
\sum_{i}V_ip_i&=&2 I_p + E_p - 2 D_p,
\end{eqnarray}
and
\begin{eqnarray}
L&=&C + I_p + I_n -\sum_{i}V_i,\\
R&=&\sum_{i}V_i-1,
\end{eqnarray}
we get the master formula\footnote{For $D_p=0$ one can easily verify that the
equations~(\ref{PowerWein}) and (\ref{powerc}) are identical.}
\begin{eqnarray}
\label{PowerWein}
\nu&=&4-3N-D_p-E_p+\sum_{i}V_i\kappa_i,
\end{eqnarray}
where
\begin{eqnarray}
\label{Kapa}
\kappa_i&=&d_i + \frac{3}{2}n_i+p_i-4,
\end{eqnarray}
and $n_i$ denotes the number of nucleons in vertex $i$, which can only be even,
and $p_i$ counts the number of pion fields associated to the vertex $i$.

In the case of the nucleon-nucleon interaction there are no external pions so
that $D_p=E_p=0$ and we get a simplified formula for the chiral order
\begin{eqnarray}
\nu&=&4-3N+\sum_{i}V_i\kappa_i.
\end{eqnarray}
Because of chiral symmetry, $\kappa_i$ has to be a positive
number, thus the order of every diagram is bounded from below. The lowest
possible order for a $\hat{Q}$-box is $\nu_{\textrm{\tiny min}}=6-3N$ 
and is represented
through the lowest contact term from ${\cal L}_{NN}^{(0)}$ or 
the pion exchange diagram with the vertices from ${\cal L}_{\pi N}^{(1)}$.
We understand from now on the $\hat{Q}$-box as organized in a chiral expansion,
\begin{eqnarray}
\hat{Q}(E)&=&\sum_{i=0}^{\infty}\hat{Q}^{(6-3N+i)}(E),
\end{eqnarray}
where the upper index denotes the chiral order. Analogously we can write the
higher $\hat{Q}$-boxes as
\begin{eqnarray}
\hat{Q}_n(E_1,...,E_{n+1})&=&\sum_{i=0}^{\infty}
\hat{Q}_{n}^{(6-3N-n+i)}(E_1,...,E_{n+1}),
\end{eqnarray}
with
\begin{eqnarray}
\hat{Q}_{n}^{(6-3N-n+i)}(E_1,...,E_{n+1})&=&\sum_{k=1}^{n+1}C_k(E_1,...,E_{n+1})
\hat{Q}^{(6-3N+i)}(E_k),
\end{eqnarray}
where $C_k(E_1,...,E_{n+1})$ is defined in Eq.~$(\ref{CKonstK})$.
The following  combination of $\hat{Q}$-boxes
\begin{eqnarray}
\hat{O}&=&\hat{Q}_{n}^{(6-3N-n+i)}(E_{1,1},...,E_{1,n+1})
\hat{Q}_{m}^{(6-3N-m+j)}(E_{2,1},...,E_{2,m+1})\nonumber\\
&=&\hat{Q}_{n}^{(6-3N-n+i)}(E_{1,1},...,E_{1,n+1})
|\alpha\rangle\langle\alpha|
\hat{Q}_{m}^{(6-3N-m+j)}(E_{2,1},...,E_{2,m+1})\nonumber
\end{eqnarray}
leads to the chiral order $\nu(\hat{O})$:
\begin{eqnarray}
\nu (\hat{O}) &=& 6 - 3N - n + i + 6 - 3N - m + j + 3(N-1)\nonumber\\
&=& 9 - 3N + i + j - n - m,
\end{eqnarray}
where the factor $3(N-1)$ comes from the integration over all relative 
coordinates of the intermediate state $|\alpha\rangle\langle\alpha|$. 
For a general
combination of $\hat{Q}$-boxes
\begin{eqnarray}
\hat{O}&=&\hat{Q}_{n_1}^{(6-3N-n_1+i_1)}(E_{1,1},...,E_{1,n_1+1})\cdots
\hat{Q}_{n_m}^{(6-3N-n_m+i_m)}(E_{m,1},...,E_{m,n_m+1})\nonumber
\end{eqnarray}
we get
\begin{eqnarray}
\label{QboxComb}
\nu(\hat{O})&=&m(6-3N) - (n_1 + ... + n_m) + (i_1 + ... + i_m) + (m - 1)3(N-1)
\nonumber\\
&=& 3m + 3 - 3N - (n_1 + ... + n_m) + (i_1 + ... + i_m).
\end{eqnarray}
Now we are ready to proof that one can neglect higher $\hat{Q}$-boxes
for the case of pure nucleonic
scattering. At the first sight it is not clear, why the $\hat{Q}$-box 
contributions with increasing $\kappa$ number should increase the chiral order.
Conversely, we see from the last
equation that the order $\nu(\hat{O})$ decreases with increasing
$\kappa=n_1 + \cdots + n_m$. To clarify this we first point out that 
the steps for
the derivation of the operators $R$ and $A^\dagger A$ in the 
degenerate
and in the non-degenerate case were the same. The coefficients in front of the
$\hat{Q}$-boxes are the same, the difference is only in the energy-structure
of the $\hat{Q}$-boxes. From the degenerate case we learned that the 
coefficients $f(n_1,...,n_m)$ and $h(n_1,...,n_m)$ of the operators
$R$ and $A^\dagger A$ restrict
the numbers $n_1,...,n_m$, namely $f(n_1,...,n_m)\ne0$ only if the condition
$n_1+...+n_m=m-1$ is valid and $h(n_1,...,n_m)\ne0$ only if the condition
$n_1+...+n_m=m$ is valid. As a result we get for the operator $R$ that
\begin{eqnarray}
n_1+...+n_m&=& m-1
\end{eqnarray}
and from Eq.~(\ref{QboxComb})
\begin{eqnarray}
\nu(\hat{O})&=&3m + 3 - 3N - m + 1 + (i_1+...+i_m)\nonumber\\
&=&2m+4-3N+(i_1+...+i_m).
\nonumber
\end{eqnarray} 
The NNNLO will be reached when $\nu=10-3N$. One obtains from the above
equation that $m\le3$ and $\kappa=n_1+...+n_m\le2$. All the other possibilities 
are beyond  NNNLO. For the operator $A^\dagger A$ we get 
\begin{eqnarray}
n_1+...+n_m&=&m.\nonumber
\end{eqnarray}
However, we have to remember that the operator $A^\dagger A$ appears
in the hermitean potential $W$ only in combination with the operator $R$,
which can be directly seen from Eq.~$(\ref{WPotGeneral})$. For this
reason a $\hat{Q}$-box combination $\hat{Q}_{n_1}\cdots\hat{Q}_{n_m}$ from
$A^\dagger A$ has to be multiplied by the $\hat{Q}$-box combination
$\hat{Q}_{l_1}\cdots\hat{Q}_{l_k}$ from $R$. So for the whole operator we
get $n_1+...+n_m=m$ and $l_1+...+l_k=k-1$ and altogether 
\begin{eqnarray}
\kappa=n_1+...+n_m+l_1+...+l_k=m+k-1.
\end{eqnarray}
From the above considerations we see, that the condition $m+k\le3$ has to
be valid. 
With similar arguments one can show that the operators of the form
\beq
\left(A^\dagger A\right)^m R\left(A^\dagger A\right)^n
\eeq 
with $m+n>2$ contribute to orders  higher than $10-3N$.

\section{Power counting for pion scattering on nucleons}
\label{app:powerc}
In this section we want to consider the scattering of pions
off nucleons, so we define our model-space as a space, which contains the
states with only nucleons and nucleon states with at least one pion or
delta. The states with more
than two pions are in the complementary space. The power counting of
Weinberg given in Eq.~(\ref{PowerWein}) remains the same but 
$D_p$ and $E_p$ do not vanish any more, because external pions can appear.

\begin{figure}
\center
\includegraphics[height=4cm]{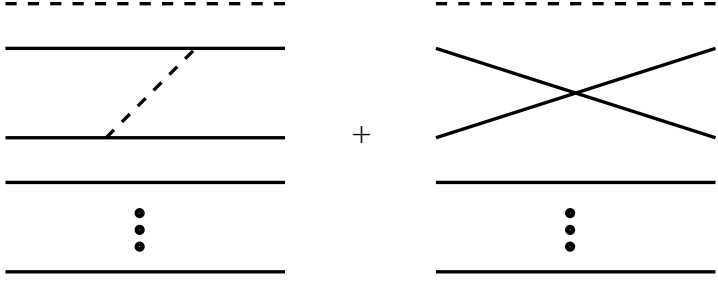}
\caption{$P^1\hat{Q}(E)P^1$-boxes to leading order.\label{lowest} } 
\end{figure}

\begin{figure}
\center
\includegraphics[height=4cm]{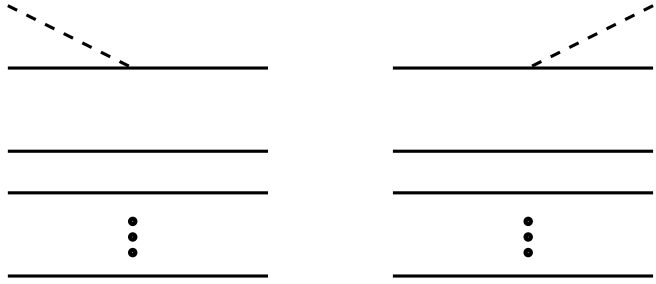}
\caption{$P^1\hat{Q}(E)P^0$ and $P^0\hat{Q}(E)P^1$-boxes to leading 
order.\label{lowestNpi} } 
\end{figure}

We have again to prove the assumption that one can neglect the higher $\hat{Q}$-boxes
for this kind of processes. For this reason let us introduce a convenient
notation: we write the projection operator $P$ as $P=P^0+P^1$, where $P^0$
and $P^1$ project to the pure nucleonic states and nucleonic states with
one pion. Utilizing this notation one can write a $\hat{Q}$-box 
as a two--by--two matrix:
\begin{eqnarray}
\left(
\begin{array}{cc}
P^1 \hat{Q}(E) P^1 & P^1 \hat{Q}(E) P^0\\
P^0 \hat{Q}(E) P^1 & P^0 \hat{Q}(E) P^0
\end{array}
\right)~.
\end{eqnarray}
The multiplication of $\hat{Q}$-boxes is nothing but
 a matrix multiplication.
At this stage there appears one problem: in the power counting 
Eq.~$(\ref{powerc})$ we did not
count the phase-space factors of external pions. However, when we construct a 
$\hat{Q}$-box we a priori do not know whether the outgoing pions of the box
are internal or external. Indeed the assumption that all outgoing pion lines 
of a $\hat{Q}$-box are external results in a contradiction, because by 
multiplication of two $\hat{Q}$-boxes external pion lines can become internal.
For this reason let us introduce some kind of internal power counting, where
we count all phase-space factors of pions, and denote the internal chiral order as 
$\tilde{\nu}$. The number of phase-space factors of external pions is given
by $E_p-2D_p$. We get the internal order by simply subtracting the factor
$(E_p/2-D_p)$ from the original order:
\begin{eqnarray}
\tilde{\nu}&=&\nu - \frac{E_p}{2}+D_p\,=\,4-3N-\frac{3}{2}E_p+\sum_{i}V_i\kappa_i 
\end{eqnarray}
Every matrix-element has its definite minimal original and internal order. 
The minimal internal order of 
 $P^1\hat{Q}(E) P^1$ is $\tilde{\nu}_{\textrm{\tiny min}}=3-3N$ 
and is represented by the
diagrams of Figure~\ref{lowest}. The lowest internal orders of elements
$P^0\hat{Q}(E) P^1$ and $P^1\hat{Q}(E) P^0$ is
$\tilde{\nu}_{\textrm{\tiny min}}=3-3N+1/2$ and
is represented by the diagrams of Figure~\ref{lowestNpi}. Finally the lowest
internal order of elements $P^0 \hat{Q}(E)P^0$ is given by 
$\tilde{\nu}_{\textrm{\tiny min}}=6-3N$ and is represented through
the contact interaction. All the vertices in the above mentioned
diagrams stem from ${\cal L}_{\pi N}^{(1)}$ and 
${\cal L}_{N N}^{(0)}$. To have an overview let us define an order matrix
as
\begin{eqnarray}
\tilde{\nu}\left(
\begin{array}{cc}
P^1 \hat{Q}(E) P^1 & P^1 \hat{Q}(E) P^0\\
P^0 \hat{Q}(E) P^1 & P^0 \hat{Q}(E) P^0
\end{array}
\right)&=&
\left(
\begin{array}{cc}
\tilde{\nu}(P^1 \hat{Q}(E) P^1) & \tilde{\nu}(P^1 \hat{Q}(E) P^0)\\
\tilde{\nu}(P^0 \hat{Q}(E) P^1) & \tilde{\nu}(P^0 \hat{Q}(E) P^0)
\end{array}
\right).
\end{eqnarray}
With this notation we have for the minimal internal chiral order of 
$\hat{Q}$-boxes
\begin{eqnarray}
\tilde{\nu}_{\textrm{\tiny min}}\left(
\begin{array}{cc}
P^1 \hat{Q}(E) P^1 & P^1 \hat{Q}(E) P^0\\
P^0 \hat{Q}(E) P^1 & P^0 \hat{Q}(E) P^0
\end{array}
\right)&=&
\left(
\begin{array}{cc}
3-3N & 3-3N+\frac{1}{2}\\
3-3N+\frac{1}{2} & 6-3N
\end{array}
\right)
\end{eqnarray}
and for higher $\hat{Q}$-boxes
\begin{eqnarray}
\label{Q1Min}
\hat{O}&=&\left(
\begin{array}{cc}
P^1 \hat{Q}_n(E_1,...,E_{n+1}) P^1 & P^1 \hat{Q}_n(E_1,...,E_{n+1})P^0\\
P^0 \hat{Q}_n(E_1,...,E_{n+1}) P^1 & P^0 \hat{Q}_n(E_1,...,E_{n+1})P^0
\end{array}
\right)\nonumber\\
\tilde{\nu}_{\textrm{\tiny min}}(\hat{O})&=&
\left(
\begin{array}{cc}
3-3N-n & 5-3N+\frac{1}{2}-n\\
5-3N+\frac{1}{2}-n & 8-3N-n
\end{array}
\right),
\end{eqnarray}
where $n\ge1$. 
The maximal order, which corresponds to order $q^4$ in pion scattering off nucleons 
(standard single-nucleon counting) is given by $\nu_{\textrm{\tiny max}}=7-3N$.
The following matrix of $\hat{Q}$-boxes
\begin{eqnarray}
\left(
\begin{array}{cc}
P^1 \hat{Q}_{n}^{(3-3N-n+j)}(E_{1},...,E_{n+1}) P^1 & 
P^1 \hat{Q}_n^{(3-3N+\frac{1}{2}-n+j)}(E_{1},...,E_{n+1}) P^0\\
P^0 \hat{Q}_n^{(3-3N+\frac{1}{2}-n+j)}(E_{1},...,E_{n+1}) P^1 & 
P^0 \hat{Q}_n^{(6-3N-n+j)}(E_{1},...,E_{n+1}) P^0
\end{array}
\right)\nonumber
\end{eqnarray}
can also be represented as
\begin{eqnarray}
P^{i_1}\hat{Q}_n^{(5-\frac{3}{2}(i_1+i_2)+\delta_{i_1,i_2}-3N-n+j)}
(E_1,...,E_{n+1})P^{i_2}.\nonumber
\end{eqnarray}
For the order of $m$ multiplied $\hat{Q}$-boxes
\begin{eqnarray}
\hat{O}&=&P^{i_1}\hat{Q}_{n_1}^{(5-\frac{3}{2}(i_1+i_2)
+\delta_{i_1,i_2}-3N-n_1+j_1)}
(E_{1,1},...,E_{1,n_1+1})P^{i_2}\nonumber\\
&...&P^{i_m}\hat{Q}_{n_m}^{(5-\frac{3}{2}(i_m+i_{m+1})
+\delta_{i_m,i_{m+1}}-3N-n_m+j_m)}
(E_{m,1},...,E_{m,n_m+1})P^{i_{m+1}}\nonumber
\end{eqnarray}
we get
\begin{eqnarray}
\tilde{\nu}(\hat{O})&=&m(5-3N)-(n_1+...+n_m)+(j_1+...+j_m)+(m-1)3N\\
&-&\frac{3}{2}(i_1+i_{m+1})+\sum_{k=1}^{m}\delta_{i_k,i_{k+1}}-3(i_2+...+i_m)
-3\sum_{k=2}^{m}\delta_{i_k,0}.\nonumber
\end{eqnarray}
The last sum of Kronecker-symbols comes from the integration over the
intermediate states. The integration over
the relative coordinates brings in the case of a pure nucleonic 
intermediate state $3(N-1)$
 and in the case of a state with one pion $3N$ extra dimensions in 
the power counting. For the
construction of the operator $R$ we can use the
following relations 
\begin{eqnarray}
m-1&=&n_1+.. +.n_m,\nonumber\\
m-1&=&i_2+...+i_m+\sum_{k=2}^{m}\delta_{i_k,0}\nonumber,
\end{eqnarray}
where the fact that $i_2,...,i_m\in\left\{0,1\right\}$ has been used, and we get
for the internal order
\begin{eqnarray}
\tilde{\nu}(\hat{O})&=&m-3N+4+\left(\sum_{k=1}^{m}\delta_{i_k,i_{k+1}}
-\frac{3}{2}i_1-\frac{3}{2}i_{m+1}\right)+(j_1+...+j_m)\\
&\ge& m-3N+1\nonumber,
\end{eqnarray}
such that the relation $m\le 6$ has to hold. However, the value of $m$ can be
further decreased. The above inequality becomes an equality, if $i_1=i_{m+1}=1$
 and the sum of the Kronecker-symbols is zero. Let us discuss first
this elastic case. For every $k=1,...,m$ the condition $i_k\ne i_{k+1}$ is valid.
In this case at least one pure nucleonic intermediate state has to appear.
It follows that $D_p=0$ and the original order of this combination of 
$\hat{Q}$-boxes is $\nu=\tilde\nu+E_p/2=m-3N+2$ leading  to the condition $m\le 5$.
In the inelastic
case $i_1=0$ or $i_{m+1}=0$ the original order is given by $\nu\ge m-3N+3$,
such that even the condition $m\le4$ has to be valid. Further 
decreasing of $m$ will be obtained in the following manner: 
let us assume the equality
$m=5$. In the elastic case $i_1=i_{m+1}=1$ the condition $\sum_{k=1}^{m}
\delta_{i_k,i_{k+1}}\le1$ has to be valid. For $m=5$  only the
equality is possible, so that there appears at least one pure nucleonic
intermediate state and the original order increases by one, such that
$\nu=8-3N$ and is beyond the fourth order. In the inelastic case 
the condition $m=5$ is excluded and we get
$m\le4$. Let us assume that $m=4$. In the elastic case
we get the condition $\sum_{k=1}^{m}\delta_{i_k,i_{k+1}}\le2$.
In any case one pure intermediate nucleonic state has to appear, otherwise
we would get $\sum_{k=1}^{m}\delta_{i_k,i_{k+1}}=4$. This restricts the
above estimation to $\sum_{k=1}^{m}\delta_{i_k,i_{k+1}}\le1$. For $m=4$ only
the equality $\sum_{k=1}^{m}\delta_{i_k,i_{k+1}}=0$ is possible with
the $\hat{Q}$-box combination given by 
$P^1\hat{Q}_{n_1}P^0\hat{Q}_{n_2}P^1\hat{Q}_{n_3}P^0\hat{Q}_{n_4}P^1$. However,
 for $m=4$ at least one higher $\hat{Q}$-box has to appear. For the above 
combination of $\hat{Q}$-boxes one gets an increase of the original order
by two, such that $\nu=8-3N$. In the inelastic case  we
get the condition $\sum_{k=1}^{m}\delta_{i_k,i_{k+1}}=0$, which is impossible
for $m=4$. This proves that the $\hat{Q}$-box contributions to $R$ 
with $\kappa>2$ are of higher order than $7-3N$, that is they start to
contribute at ${\cal O}(q^5)$ in the standard single--nucleon counting.
The proof that the operators
\beq
\left(A^\dagger A\right)^m R\left(A^\dagger A\right)^n
\eeq
are of higher order  than $7-3N$ if  $m+n>2$  can be given in the similar manner 
(the same holds for the $\kappa >2$ contributions if $m+n\le 2$).

\pagebreak

\end{document}